\documentclass[aps,prb,twocolumn,showpacs,floatfix]{revtex4}%
\usepackage{amsmath,amsfonts,amssymb}

\usepackage{graphicx,color}
\definecolor{darkgreen}{rgb}{0,0.5,0}
\definecolor{darkblue}{rgb}{0,0,0.2}
\definecolor{purple}{rgb}{0.35,0,0.35}
\definecolor{orange}{rgb}{1,0.5,0}

\usepackage{color,fancyhdr}
\usepackage[english]{babel}
\RequirePackage[hyperindex,colorlinks,bookmarksnumbered,plainpages=true]{hyperref}
\hypersetup{linkcolor=blue,urlcolor=darkblue,citecolor=darkgreen}
\usepackage{hyperref}

\def\ie{{\it i.e. }}
\def\eg{{\it e.g. }}
\def\cf{{\it cf.\ }}
\def\lhs{{\it l.h.s. }}

\def\wrt{{\it w.r.t. }}

\newcommand{\Eq}[1]{Eq.~(\ref{#1})}

\newcommand{\Fig}[1]{Fig.~\ref{#1}}

\newcommand{\NC}{{N_{\mathrm{c}}}}
\newcommand{\pdag}{{\phantom{\dagger}}}

\newcommand{\interact}{\textrm{int}}

 \newcommand{\IRLM}{\textrm{IRLM}}
\newcommand{\SIAM}{\textrm{SIAM}}

\def\I{{\rm I}}  
\def\F{{\rm F}}  
\def\X{{X}}      
\def\x{P}        
\def\G{{\rm G}}  

\newcommand{\dd}{{\mathrm{d}}}
\newcommand{\Hb}{{\hat{H}_{\rm b}}}

\newcommand{\HI}{{\hat{H}_\I}}
\newcommand{\HF}{{\hat{H}_\F}}
\newcommand{\HX}{{\hat{H}_\X}}

\newcommand{\HlocX}{\hat H_{\dd,\X}}
\newcommand{\Hint}{\hat{H}_{\interact}}
\newcommand{\elocX}{\varepsilon_{d\mu,\X}}

\newcommand{\Ef}{\varepsilon^f}
\newcommand{\Vlarge}{V_{\rm large}}

\newcommand{\dao}{\Delta_{\mathrm{AO}}}
\newcommand{\daomu}{\Delta_{\mathrm{AO},\mu}}
\newcommand{\dcharge}{\Delta_{\mathrm{ch}}}
\newcommand{\dchargemu}{\Delta_{\mathrm{ch},\mu}}
\newcommand{\dphase}{\Delta_{\mathrm{ph}}}
\newcommand{\sea}{{\mathrm{sea}}}
\newcommand{\imp}{{\mathrm{dot}}} 
\newcommand{\dimp}{{\Delta_\imp}}
\newcommand{\dsea}{{\Delta_\sea}}
\newcommand{\charge}{{\mathrm{ch}}}

\newcommand{\edf}{\varepsilon_{\dd, \F}}

\newcommand{\tot}{{\mathrm{tot}}}
\newcommand{\even}{{\mathrm{even}}}
\newcommand{\odd}{{\mathrm{odd}}}

\begin{document}
\title{Anderson Orthogonality
and the Numerical Renormalization Group}

\author{Andreas \surname{Weichselbaum}}
\affiliation{
  Physics Department, Arnold Sommerfeld Center for Theoretical
  Physics, and Center for NanoScience,
  Ludwig-Maximilians-Universit\"at, 80333 Munich,
  Germany
}

\author{ Wolfgang \surname{M{\"u}nder}}
\affiliation{
  Physics Department, Arnold Sommerfeld Center for Theoretical
  Physics, and Center for NanoScience,
  Ludwig-Maximilians-Universit\"at, 80333 Munich,
  Germany
}

\author{Jan \surname{von Delft}}
\affiliation{
  Physics Department, Arnold Sommerfeld Center for Theoretical
  Physics, and Center for NanoScience,
  Ludwig-Maximilians-Universit\"at, 80333 Munich,
  Germany
}

\begin{abstract}

Anderson Orthogonality (AO) refers to the fact that the ground states
of two Fermi seas that experience different local scattering
potentials, say $| G_\I \rangle$ and $| G_\F\rangle$, become
orthogonal in the thermodynamic limit of large particle number $N$,
in that $|\langle G_\I| G_\F \rangle| \sim N^{-\frac{1}{2} \dao^2}$
for $N \to \infty$. We show that the numerical renormalization group
offers a simple and precise way to calculate the exponent $\dao$: the
overlap, calculated as function of Wilson chain length $k$, decays
exponentially, $\sim e^{-k \alpha}$, and $\dao$ can be extracted
directly from the exponent $\alpha$. The results for $\dao$ so
obtained are consistent (with relative errors typically smaller than
1\%) with two other related quantities that compare how ground state
properties  change upon switching from $|\G_\I \rangle$ to
$|\G_\F\rangle$: the difference in scattering phase shifts at the
Fermi energy, and the displaced charge flowing in from infinity. We
illustrate this for several nontrivial interacting models, including
systems that exhibit population switching.

\end{abstract}
\date{\today}
\maketitle

  \section{Introduction}

In 1967, Anderson considered the response of a Fermi sea to a change
in local scattering potential and made the following observation:
\cite{Anderson1967} the ground states $| G_\I \rangle$ and $| G_\F
\rangle$ of the Hamiltonians $\HI$ and $\HF$ describing the system
before and after the change, respectively, become orthogonal in the
thermodynamic limit, decaying with total particle number $N$ as
\begin{eqnarray}
|\langle G_\I | G_\F \rangle|
  \sim N^{-\frac{1}{2} \dao^2} \; ,
\label{eq:psi12}
\end{eqnarray}
because the single-particle states comprising the two Fermi seas are
characterized by different phase shifts.

Whenever the Anderson orthogonality (AO) exponent $\dao$ is finite,
the overlap of the two ground state wave functions goes to zero as
the system size becomes macroscopic. As a consequence, matrix
elements of the form $|\langle G_{\I} | \hat { \mathcal O} | G_{\F}
\rangle |$, where $\hat {\mathcal O}$ is a local operator acting at
the site of the localized potential, necessarily also vanish in the
thermodynamic limit. This fact has far-reaching consequences,
underlying several fundamental phenomena in condensed matter physics
involving quantum impurity models, \ie models describing a Fermi
sea coupled to localized quantum degrees of freedom. Examples are the
Mahan exciton (ME) and the Fermi-edge singularity (FES)
\cite{Mahan67,Schotte1969,Schotte1969a,Nozieres1969} in absorption
spectra, and the Kondo effect \cite{Yuval1970} arising in magnetic
alloys \cite{Kondo1964} or in transport through quantum dots.
\cite{Goldhaber-Gordon1998PRL} For all of these, the
low-temperature dynamics is governed by the response of the Fermi sea
to a sudden switch of a local scattering potential. More recently,
there has also been growing interest in inducing such a sudden
switch, or quantum quench, by optical excitations of a quantum dot
tunnel-coupled to a Fermi sea, in which case the post-quench dynamics
leaves fingerprints, characteristic of AO, in the optical absorption
or emission line shape. \cite{Helmes05,Tureci2011,Latta2011}

The intrinsic connection of local quantum quenches to the scaling of
the Anderson orthogonality with system size can be intuitively
understood as follows. Consider an instantaneous event at the
location of the impurity at time $t=0$ in a system initially in
equilibrium. This local perturbation will spread out spatially, such
that for $t>0$, the initial wave function is affected only within a
radius $L\simeq v_f t$ of the impurity, with $v_f$ the Fermi
velocity. The AO finite-size scaling in \Eq{eq:psi12} therefore
directly resembles the actual experimental situation, and in
particular allows the exponent $\dao$ to be directly related to the
exponents seen in experimental observables at long time scales, or at
the threshold frequency in Fourier space.\cite{Muender2011b}

A powerful numerical tool for studying quantum impurity models is the
numerical renormalization group (NRG),\cite{Wilson1975,Bulla2008}
which allows numerous static and dynamical quantities to be calculated
explicitly, also in the thermodynamic limit of infinite bath
size.  The purpose of the present paper is to point out that NRG
also offers a completely straightforward way to calculate the overlap
$|\langle G_\I | G_\F \rangle|$ and hence to extract $\dao$.  The
advantage of using NRG for this purpose is that NRG is able to deal
with quantum impurity models that in general also involve local
\emph{interactions}, which are usually not tractable
analytically. Although Anderson himself did not include local
interactions in his considerations,\cite{Anderson1967} his prediction
(\ref{eq:psi12}) still applies, provided the ground states $|G_{\I,
  \F} \rangle$ describe Fermi liquids. This is the case for most
impurity models (but not all; the two-channel Kondo model is a notable
exception).
Another useful feature of NRG is that it allows consistency checks on
its results for overlap decays, since $\dao$ is known to be related to
a change of scattering phase shifts at the Fermi surface. These phase
shifts can be calculated independently, either from NRG energy flow
diagrams, or via Friedel's sum rule from the displaced charge, as will
be elaborated below.

A further concrete motivation for the present study is to develop a
convenient tool for calculating AO exponents for quantum dot models
that display the phenomenon of population switching.
\cite{Silvestrov2000,Golosov2006,Karrasch2007PRL,Goldstein2010}  In
such models, a quantum dot tunnel-coupled to leads contains levels of
different widths, and is capacitively coupled to a gate voltage that
shifts the levels energy relative to the Fermi level of the leads.
Under suitable conditions, an (adiabatic) sweep of the gate voltage
induces an inversion in the population of these levels (a so called
population switch), implying a change in the local potential seen by
the Fermi seas in the leads. In this paper, we verify that the method
of extracting $\dao$ from $\langle G_\I | G_\F \rangle$ works
reliably also for such models. In a separate publication
\cite{Muender2011b} we will use this method to analyze whether AO can
lead to a quantum phase transition in such models, as suggested in
Ref.~\onlinecite{Goldstein2010}.

The remainder of this paper is structured as follows: In
Sec.~\ref{Sec:AO} we define the AO exponent $\dao$ in general terms,
and explain in Sec.~\ref{Sec:NRG} how NRG can be used to calculate it.
Sec.~\ref{Sec:Results} presents numerical results for several
interacting quantum dot models of increasing complexity: first the
spinless interacting resonant level model (IRLM), then the
single-impurity Anderson model (SIAM), followed by two models
exhibiting population switching, one for spinless, the other for
spinful electrons. In all cases, our results for $\dao$ satisfy all
consistency checks to within less than 1\%.


  \section{Definition of Anderson orthogonality \label{Sec:AO}}

\subsection{AO for a single channel} \label{sec:AO-single-channel}

To set the stage, let us review AO in the context of a free Fermi sea
involving a single species or channel of noninteracting electrons
experiencing two different local scattering potentials. The initial
and final systems are described in full by the Hamiltonians $\HI$ and
$\HF$, respectively. Let $\hat{c}_{\varepsilon,\X}^\dagger |0\rangle$
be the single-particle eigenstates of $\HX$ characterized by the
scattering phase shifts $\delta_X (\varepsilon)$, where $\X \in
\{\I,\F\}$ and $\hat{c}_{\varepsilon,\X}^\dagger$ are fermion
creation operators, and let $\Ef$ be the same Fermi energy
for both Fermi seas $| G_{\X}\rangle$. Anderson showed that in the
thermodynamic limit of large particle number, $N \to \infty$, the
overlap
\begin{equation}
\left\langle G_{\I}| G_{\F}\right\rangle = \left\langle 0\right\vert
\prod_{\varepsilon<\Ef}\hat{c}_{\varepsilon,\I}^{\ } \cdot
\prod_{\varepsilon<\Ef}\hat{c}_{\varepsilon,\F}^{\dagger}\left\vert 0\right\rangle
\label{eq:overlap}
\end{equation}
decays as in \Eq{eq:psi12}, \cite{Anderson1967,Schotte1969a}
where $\dao$ is equal to the difference in single-particle phase
shifts at the Fermi level,
\begin{equation}
   \dao = \dphase \equiv
   [ \delta_{\F}(\Ef) - \delta_{\I}(\Ef)] / \pi \; .
   \label{eq:AO-phaseshift}
\end{equation}
The relative sign between $\dao$ and $\dphase$ ($+$, not $-$) does
not affect the orthogonality exponent $\dao^2$, but follows standard
convention [Ref.~\onlinecite{Friedel1956}, Eq.~(7), or
Ref.~\onlinecite{Langreth1966}, Eq.~(21)].

In this paper we will compare three independent ways of calculating
$\dao$.
(i) The first approach calculates the overlap $|\langle G_\I|G_\F
\rangle|$ of \Eq{eq:psi12} explicitly as a function of (effective)
system size.  The main novelty of the present paper is to point out
that this can easily be done in the framework of NRG, as will be
explained in detail in section~\ref{Sec:NRG}.

(ii) The second approach is to directly calculate $\dphase$ via
\Eq{eq:AO-phaseshift}, since the extraction of phase shifts
$\delta_\X (\Ef)$ from NRG finite-size spectra is
well-known: \cite{Wilson1975} provided that $\HX$ describes a Fermi
liquid, the (suitably normalized) fixed point spectrum of NRG can be
reconstructed in terms of equidistant free-particle levels shifted by
an amount determined by  $\delta_\X(\Ef)$. The many-body
excitation energy of an additional particle, a hole and a
particle-hole pair thus allow the phase shift $\delta_\X
(\Ef)$ to be determined unambiguously.

(iii) The third approach exploits Friedel's sum
rule,\cite{Friedel1956} which  relates the difference in phase shifts
to the so called \emph{displaced charge} $\dcharge$ via $\dcharge =
\dphase$. Here the displaced charge $\dcharge$ is defined as the
charge in units of $e$ (\ie the number of electrons) flowing inward
from infinity into a region of large but finite volume, say
$\Vlarge$, surrounding the scattering location, upon switching from
$\HI$ to $\HF$:
\begin{eqnarray}
   \dcharge &\equiv&
   \left\langle G_{\F} \right\vert \hat{n}_\tot \left\vert G_{\F} \right\rangle -
   \left\langle G_{\I} \right\vert \hat{n}_\tot \left\vert G_{\I} \right\rangle
   \nonumber \\
   &\equiv& \dsea + \dimp \; .
\label{eq:displacedcharge}
\end{eqnarray}
Here $\hat{n}_{\rm tot} \equiv \hat{n}_{\rm sea} + \hat{n}_\imp$,
where  $\hat{n}_\sea$ is the total number of Fermi
see electrons within $\Vlarge$, whereas $\hat{n}_\imp$
is the local charge of the scattering site, henceforth called
``dot''.

To summarize, we have the equalities
\begin{equation}
	\label{eq:consistency}
	\dao^2 = \dphase^2 = \dcharge^2  \;  ,
\end{equation}
where all three quantities can be calculated independently and
straightforwardly within the NRG. Thus \Eq{eq:consistency}
constitutes a strong consistency check. We will demonstrate below
that NRG results satisfy this check with good accuracy (deviations
are typically below 1\%).

\subsection{AO for multiple channels} \label{sec:AO-multiple-channel}

We will also consider models involving several independent and
conserved channels (\eg spin in spin-conserving models). In the
absence of interactions, the overall ground state wave function is
the product of those of the individual channels. With respect to AO,
this trivially implies that each channel \emph{adds} independently to
the AO exponent in \Eq{eq:psi12},
\begin{equation}
   \dao^2 = \sum_{\mu=1}^{\NC} {\dao^2}_{,\mu}
   \text{,}
   \label{eq:dao_add}
\end{equation}
where $\mu=1,\ldots,\NC$ labels the $\NC$ different channels.  We
will demonstrate below that the additive character in \Eq{eq:dao_add}
generalizes to systems with \emph{local interactions}, provided that
the particle number in each channel remains conserved. This is
remarkable, since interactions may cause the ground state wave
function to involve entanglement between local and Fermi sea degrees
of freedom from different channels. However, our results imply that
the asymptotic tails of the ground state wave function far from the
dot still factorize into a product of factors from individual
channels. In particular, we will calculate the displaced charge for
each individual channel, \cf \Eq{eq:displacedcharge},
\begin{eqnarray}
  \Delta_{\charge, \mu} &\equiv&
   \left\langle G_{\F} \right\vert \hat{n}_{\tot,\mu}
\left\vert G_{\F} \right\rangle -
   \left\langle G_{\I} \right\vert \hat{n}_{\tot,\mu}
\left\vert G_{\I} \right\rangle
   \nonumber \\
  \label{eq:displacedcharge-mu}
   &\equiv& {\dsea}_{,\mu} + {\dimp}_{,\mu} \; ,
\end{eqnarray}
where $\hat n_{\tot, \mu} = \hat n_{\sea, \mu} + \hat n_{\imp, \mu}$.
Assuming no interactions in the respective Fermi seas, it follows
from Friedel's sum rule that ${\dao^2}_{,\mu} = {\dcharge^2}_{,\mu}$,
and therefore
\begin{equation}
  \dao^2 = \sum_{\mu=1}^{\NC} {\dcharge^2}_{,\mu} \equiv \dcharge^{2} \,  ,
\label{eq:dcharge_add}
\end{equation}
where $\dcharge^2$ is the total sum of the squares of the displaced
charges of the separate channels. Equation~(\ref{eq:dcharge_add})
holds with great numerical accuracy, too, as will be shown below.

  \section{Treating Anderson Orthogonality using NRG \label{Sec:NRG}}

\subsection{General impurity models}

The problem of a noninteracting Fermi sea in the presence of a local
scatterer belongs to the general class of quantum impurity models
treatable by Wilson's NRG.\cite{Wilson1975} Our proposed approach for
calculating $\dao$ applies to \emph{any} impurity model treatable by
NRG. To be specific, however, we will focus here on generalized
Anderson impurity type models.  They describe $\NC$ different (and
conserved) species or channels of fermions that hybridize with local
degrees of freedom at the dot,
while all interaction terms are local.

We take both the initial and final ($X \in \{ \I, \F \}$)
Hamiltonians to have the generic form $\hat{H}_X = \Hb + \HlocX +
\Hint$. The first term,
\begin{eqnarray}
  \label{eq:freeFermiSeas}
  \Hb = \sum_{\mu = 1}^{\NC}
\sum_\varepsilon \varepsilon\,
\hat{c}^\dagger_{\varepsilon \mu} \hat{c}^\pdag_{\varepsilon \mu}
\text{,}
\end{eqnarray}
describes a noninteracting Fermi sea involving $\NC$ channels. ($\NC$
includes the spin index, if present.) For simplicity, we assume
a constant density of states $\rho_\mu(\varepsilon) =
\rho_{0,\mu} \theta (D - |\varepsilon|)$ for each channel with
half-bandwidth $D$.
Moreover, when representing numerical results, energies will be
measured in units of half-bandwidth, hence $D:=1$.
The Fermi sea is assumed to couple to the dot only via the local
operators $\hat{f}^\pdag_{0\mu} = \tfrac{1}{\sqrt{N_{\rm b}}}
\sum_\epsilon \hat{c}^\pdag_{\epsilon \mu} $ and $\hat{f} ^\dag
_{0\mu}$, that, respectively, annihilate or create a Fermi sea
electron of channel $\mu$ at the position of the dot, $\vec r =
0$, with a proper normalization constant $N_{\rm b}$ to ensure
$[f_{0\mu}, f^\dagger_{0\mu'}] = \delta_{\mu \mu'}$.

The second term, $\HlocX$, contains the non-interacting local
part of the Hamiltonian, including the dot-lead hybridization,
\begin{eqnarray}
\HlocX &=& \sum_{\mu=1}^{\NC} \elocX \hat n_{d \mu}
  + \sum_{\mu=1}^{\NC} \sqrt{\tfrac{2\Gamma_\mu}{\pi}}
    \bigl[ \hat{d}_{\mu}^{\dag}\hat{f}_{0\mu}^{\pdag} +
    h.c. \bigr] \text{ .}
\label{eq:Hd_general}
\end{eqnarray}
Here  $\elocX$ is the energy of dot level $\mu$ in the initial or
final configuration, and $\hat{n}_{d\mu} =  \hat{d}_{\mu}^{\dagger}
\hat{d}_{\mu}$ is its electron number. $\Gamma_\mu \equiv \pi\rho_\mu
V_\mu^2$ is the effective width of level $\mu$ induced by its
hybridization with channel $\mu$ of the Fermi sea, with $V_\mu$ the
$\mu$-conserving matrix element connecting the d-level with the bath
states $\hat{c}_{\varepsilon\mu}$, taken independent of energy, for
simplicity.

Finally, the interacting third term is given in the case of the
single-impurity Anderson model (SIAM) by the uniform Coulomb
interaction $U$ at the impurity,
\begin{eqnarray}
    \hat H^\SIAM_\interact
    &=& \tfrac{1}{2} U \hat{n}_\dd (\hat{n}_\dd - 1) \text{,}
\label{eq:Hint-SIAM}
\end{eqnarray}
with $\hat n_\dd = \sum_\mu \hat n_{d\mu}$, while in case of the
interacting resonant level model (IRLM), the interacting part is
given by
\begin{eqnarray}
    \hat H^\IRLM_\interact & = &   U' \hat n_\dd \hat n_0 \text{ ,}
\label{eq:Hint-RLM}
\end{eqnarray}
with $\hat n_0 = \sum_\mu f^\dagger_{0,\mu}
f^{\phantom{\dagger}}_{0,\mu} \equiv \sum_\mu\hat n_{0,\mu}$.
In particular, most of our results are for the one- or two-lead
versions of the SIAM for spinful or spinless electrons,
\begin{eqnarray}
  \label{eq:SIAM}
  \hat H^\SIAM_\X = \Hb + \HlocX + \hat H_\interact^\SIAM \;  .
\end{eqnarray}
We consider either a single dot-level coupled to a single lead
(spinfull, $\NC = 2:\ \mu \in \{\uparrow, \downarrow \}$),
or a dot with two levels coupled separately to two leads
(spinless, $\NC = 2:\ \mu \in \{1,2\}$;
spinfull, $\NC = 4:\ \mu \in \{
   1\!\!\uparrow,1\!\!\downarrow, 2\!\!\uparrow, 2\!\!\downarrow
\}$). A splitting of the energies $\elocX$ in the spin label (if
any), will be referred to as magnetic field $B$.
We also present some results for the  IRLM, for a single channel of
spinless electrons ($\NC = 1$):
\begin{eqnarray}
  \label{eq:IRLM}
  \hat H^\IRLM_\X = \Hb + \HlocX + \hat H_\interact^\IRLM \; .
\end{eqnarray}

In this paper, we
focus on the case that $\hat H_\I$ and $\hat H_\F$ differ only in the
local level positions ($\varepsilon_{d \mu, \I} \neq \varepsilon_{d
\mu, \F}$). It is emphasized, however, that our methods are
equally applicable for differences between initial and final values
of any other parameters, including the case that the interactions are
channel-specific, \eg $\sum_{\mu \mu'} U_{\mu \mu'} \hat n_{d \mu}
\hat n_{d\mu'}$ or $\sum_{\mu \mu'} U'_{\mu \mu'} \hat n_{d \mu} \hat
n_{0\mu'}$.

\subsection{AO on Wilson chains}

Wilson discretized the spectrum of $\Hb$ on a logarithmic grid of
energies $\pm D \Lambda^{-k}$ (with $\Lambda > 1$, $k = 0, 1, 2,
\dots$), thereby obtaining exponentially high resolution of
low-energy excitations. He then mapped the impurity model onto a
semi-infinite ``Wilson tight-binding chain'' of sites $k= 0$ to
$\infty$, with the impurity degrees of freedom coupled only to site
0. To this end, he made a basis transformation from the set of sea
operators $\{\hat{c}_{\varepsilon\mu} \}$ to a new set $\{ \hat{f}_{k
\mu} \}$, chosen such that they bring $\Hb$ into the tridiagonal form
\begin{eqnarray}
  \label{eq:WilsonChainH0}
  \Hb \simeq \sum_{\mu = 1}^{\NC}  \sum_{k=1}^{\infty} t_k
  (\hat{f}^\dagger_{k \mu} \hat{f}^{\pdag}_{k-1 ,\mu} + {\rm h.c.}) \; .
\end{eqnarray}
The hopping matrix elements $t_k \propto D \Lambda^{-k/2}$ decrease
exponentially with site index $k$ along the chain. Because of this
separation of energy scales, the Hamiltonian can be diagonalized
iteratively by solving a Wilson chain of length $k$ (restricting the
sum in \Eq{eq:WilsonChainH0} to the first $k$ terms) and increasing
$k$ one site at a time: starting with a short Wilson chain, a new
\emph{shell} of many-body eigenstates for a Wilson chain of length
$k$, say $|s\rangle_k$, is constructed from the states of site $k$
and the $M_K$ lowest-lying eigenstates of shell $k-1$. The latter are
the so called \emph{kept} states $|s\rangle_{k-1}^K$ of shell $k-1$,
while the remaining higher-lying states $|s\rangle_{k-1}^D$ from
that shell are \emph{discarded}.

The typical spacing between the
few lowest-lying states of shell $k$, \ie the energy scale $dE_k$,
is set by the hopping matrix element $t_k$ to the previous site,  hence
\begin{eqnarray}
  \label{eq:meanlevelspacingshellk}
  dE_k \simeq t_k \propto D \Lambda^{-k/2} \;.
\end{eqnarray}
Now, for a noninteracting Fermi sea with $N$ particles, the mean
single-particle level spacing at the Fermi energy scales as $dE
\propto D/N $. This also sets the energy scale for the mean level
spacing of the few lowest-lying many-body excitations of the Fermi
sea. Equating this to \Eq{eq:meanlevelspacingshellk}, we conclude
that a Wilson chain of length $k$ represents a Fermi sea with an
actual size $L\propto N$, \ie an \emph{effective} number of
electrons $N$, that grows \emph{exponentially} with $k$,
\begin{eqnarray}
  \label{eq:NLscaling}
\label{Escale2L}
  N \propto \Lambda^{k/2} \;.
\end{eqnarray}

Now consider two impurity models that differ only in their local
terms $\HlocX$, and let $|G_\X \rangle_k $ be the ground states of
their respective Wilson chains of length $k$, obtained via two
separate NRG runs. \cite{Helmes05} Combining Anderson's prediction
(\ref{eq:psi12}) and \Eq{eq:NLscaling}, the ground state overlap is
expected to decay exponentially with $k$ as
\begin{equation}
|{}_{k \!\!}\left\langle  G_{\I}| G_{\F}\right\rangle_k |%
\propto \Lambda^{-k \dao^{2}/4} \equiv
e^{-\alpha k} \,
\label{eq:psi12_wilson}%
\end{equation}
with
\begin{equation}
   \dao^2 = \frac{4 \alpha}{\log \Lambda} \; .
   \label{def.alpha}%
\end{equation}
Thus the AO exponent can be determined by using NRG to
\emph{directly} calculate the \lhs of
\Eq{eq:psi12_wilson} as function of chain length $k$, and extracting
$\dao$ from the exponent $\alpha$ characterizing its exponential
decay with $k$.

For \emph{noninteracting} impurity models ($U=U'=0$), a finite Wilson
chain represents a single-particle Hamiltonian for a finite number of
degrees of freedom that can readily be diagonalized numerically,
without the need for implementing NRG truncation. The ground state is
a Slater determinant of those single-particle eigenstates that are
occupied in the Fermi sea. The overlap $\langle G_\I|G_\F \rangle$ is
then given simply by the determinant of a matrix whose elements are
overlaps between the $\I$- and $\F$-versions of the occupied
single-particle states. It is easy to confirm numerically in this
manner that $\langle G_\I|G_\F \rangle \sim e^{-\alpha k}$, leading
to the expected AO in the limit $k \to \infty$. We will thus focus on
interacting models henceforth, that require the use of NRG.

In the following three subsections, we discuss several technical
aspects needed for calculating AO with NRG on Wilson chains.

\subsection{Ground state overlaps}

The calculation of state space overlaps within the
NRG is straightforward in principle\cite{Helmes05,Anders2006},
especially considering its underlying matrix product state
structure \cite{Wb07,Wb09,Schollwoeck11}. Now, the overlap in
\Eq{eq:psi12_wilson} that needs to be calculated in this
paper, is with respect to ground states as function of Wilson
chain length $k$. As such, two complications
can arise. (i) For a given $k$, the system can have several
degenerate ground states $\{|s\rangle_k^{\X}: s\in G\}$, with the
degeneracy $d_{\X,k}$ typically different for even and odd $k$. (ii)
The symmetry of the ground state space may actually differ with
alternating $k$ between certain initial and final configurations,
$\X\in\{\I,\F\}$, leading to strictly zero overlap there. A natural
way to deal with (i), is to essentially average over the degenerate
ground-state spaces, while (ii) can be ameliorated by partially
extending the ground state space to the full kept space,
$\{ |s\rangle_k^X: s \in K\}$,
as will be outlined in the following.

\begin{figure}[tb]
\begin{center}
\includegraphics[width=1\linewidth]{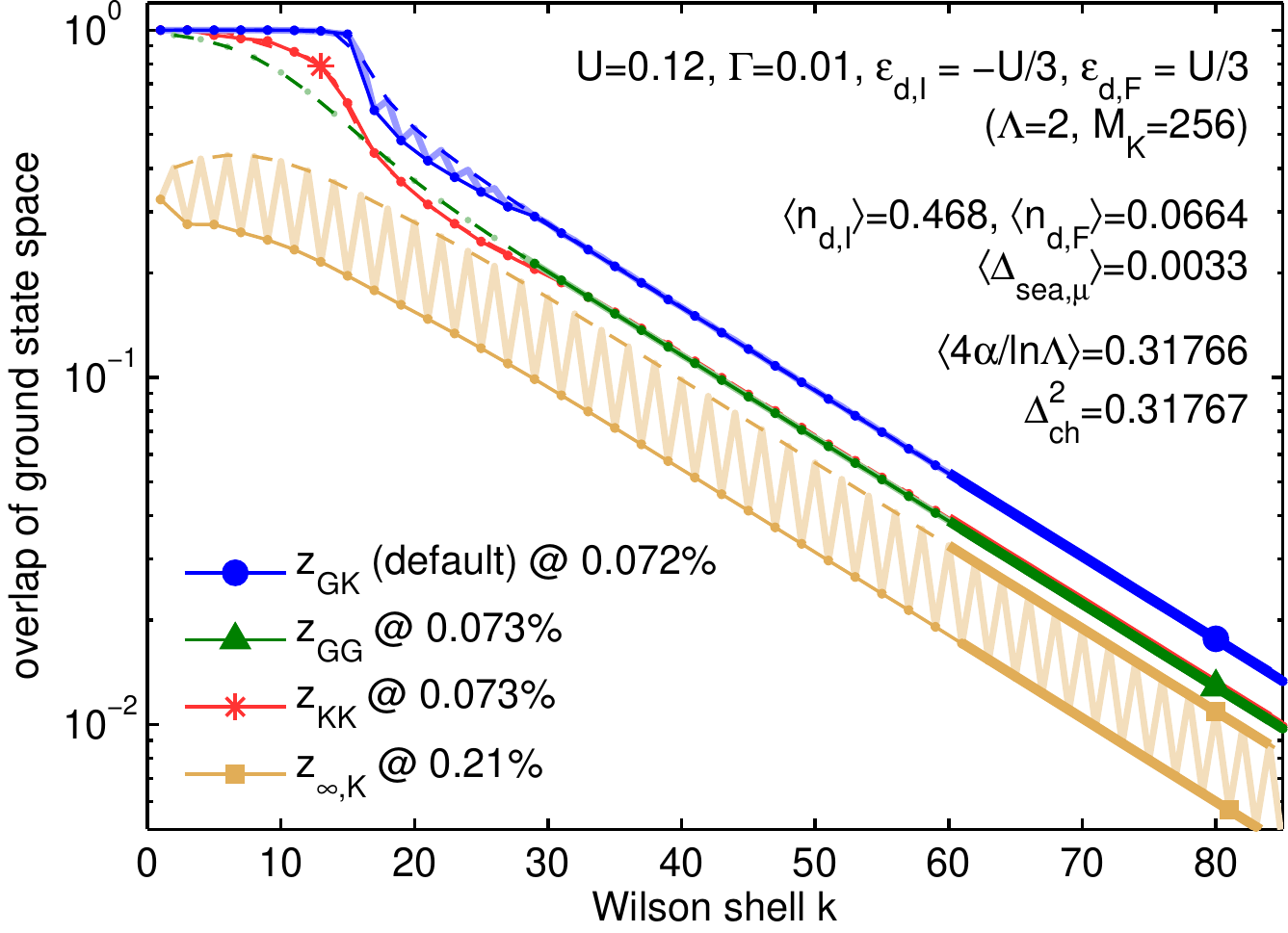}
\end{center}
\caption{
  (Color online) Anderson orthogonality for the spin-degenerate
  standard SIAM for a single lead
  [\Eq{eq:Hd_general}, $\mu \in \{\uparrow, \downarrow\}$],
  with $\mu$-independent parameters $\varepsilon_{\dd}$
  and $\Gamma$ for $\HI$ and $\HF$ as specified in the panel
  (the full $\varepsilon^\F_{\dd}$-dependence of $\dao$ for fixed
  $\varepsilon^\I_{\dd}$ is analyzed in more detail in
  \Fig{fig:SIAM}) ---
  Several alternative measures for calculating the AO-overlap
  are shown, using $z_{\x\x'}(k)$ in \Eq{eq:genoverlap}
  with $\x^{(\prime)} \in \{G,K,\infty\}$, as defined in the text.
  All overlaps are plotted for even and odd iterations separately
  to account for possible even-odd behavior within the Wilson chain
  (thin solid lines with dots, and dashed lines, respectively,
  while heavy symbols identify lines with corresponding legends).
  If even and odd data from the same $z_{\x\x'}(k)$ do not lie on
  the same smooth line, the combined data is also plotted (light
  zigzag lines) as guides to the eye. For large $k$, all AO-overlaps
  exhibit exponential decay of equal strength. Separate fits of
  $e^{\lambda - \alpha k}$ to even and odd sectors are shown
  as thick solid lines, whose lengths indicate the fitting range
  used. The values for $\dao^2$ extracted from these fits
  using \Eq{def.alpha} are in excellent agreement with the displaced
  charge $\dcharge^2$, as expected from \Eq{eq:dcharge_add}. The
  relative error is less than 1\% throughout, with the detailed
  values specified in the legend, and $\langle 4\alpha/\ln
  \Lambda\rangle$ representing the averaged value \wrt the four
  measures considered.
}
\label{fig:nrg}
\end{figure}

The $d_{\X,k}$-fold degenerate ground state subspace is described by
its projector, written in terms of the fully mixed density matrix,
\begin{eqnarray}
   \hat{\rho}_{G,k}^{\X} \equiv \tfrac{1}{d_{\X,k}}
   \sum_{s \in G}^{d_{\X,k}}\left\vert s \right\rangle \!{}_{k}^{\X}
   {}_{\, k}^{\X} \!\! \left\langle s\right\vert \text{.}
 \label{eq:def_rhoG}
\end{eqnarray}
It is then convenient to calculate the overlap of the ground
state space as follows,
\begin{eqnarray}
 z^2_{GK}(k) &\equiv&
    \mathrm{tr}_{K,k}^{\F} ( \hat{\rho}_{G,k}^{\I} ) \nonumber \\
  &=& \tfrac{1}{d_{\I,k}}
    \sum_{s \in G} \sum_{s' \in K}
    \bigl\vert
	{}_{k}^{\I} \!\! \left\langle s|s^{\prime}\right\rangle \!{}_{k}^{\F}
    \bigr\vert^2
 \text{,}
 \label{eq:def_overlap}
\end{eqnarray}
where $\mathrm{tr}_{K,k}^{\F}(\cdot)$ refers to the trace over the
kept space at iteration $k$ of the final system. The final expression
can be interpreted, up to the prefactor, as the square of the
Frobenius norm of the overlap matrix
${}_{k}^{\I}\!\!\left\langle s\vert
s^{\prime}\right\rangle \!{}_{k}^{\F}$ between the NRG states $s\in
G$ and $s^{\prime} \in K$ at iteration $k$ for the initial and
final Hamiltonian, respectively.

Note that the specific overlap in \Eq{eq:def_overlap}, as used
throughout later in this paper, not only includes the ground space of
the final system at iteration $k$, but rather includes the \emph{full
kept space} of that system. Yet each such overlap
scales as $e^{-\alpha k}$, with the same
exponent $\alpha$ for all combinations of $s$ and $s'$,
because (i) the states $|s\rangle_k^\I$ with $s\in G$ are taken
from the initial ground state space, and (ii) the states
$|s'\rangle_k^\F$ with $s \in K$
from the final kept shell differ from a final ground state
only by a small number of excitations. Therefore, Eq. (21)
is essentially equivalent, up to an irrelevant prefactor, to strictly
taking the overlap of ground state spaces as in $z^2_{GG}(k) \equiv
\mathrm{tr}_{G,k}^{\F} ( \hat{\rho}_{G,k}^{\I} )$.
This will be shown
in more detail in the following. In particular, the overlap in
\Eq{eq:def_overlap} can be easily generalized to
\begin{eqnarray}
   z^2_{\x\x'}(k) \equiv
   \mathrm{tr}_{\x',k}^{\F} ( \hat{\rho}_{\x,k}^{\I} ),
   \ \ \ 0 \le z^2_{\x\x'}(k) \le 1
\label{eq:genoverlap}
\end{eqnarray}
where $\x^{(\prime)} \in \{G,K,\infty\}$ represents the ground state
space, the full kept space, or the ground state taken at
$k\rightarrow\infty$ with respect to either initial or final system,
respectively. The overlap $z^2_{\x\x'}(k)$ in \Eq{eq:genoverlap} then
represents the fully-mixed density matrix in space $\x$ of the
initial system traced over space $\x'$ of the final system, all
evaluated at iteration $k$.

A detailed comparison for several different choices of
$z^2_{\x\x'}(k)$, including $z^2_{GG}(k)$, is provided in
Fig.~\ref{fig:nrg} for the standard SIAM with $\mu \in
\{\uparrow,\downarrow\}$). The topmost line (identified with legend
by heavy round dot) shows the overlap \Eq{eq:overlap} used as default
for calculating the overlap in the rest of the paper. This measure is
most convenient, as it reliably provides data with a smooth
$k$-dependence for large $k$, insensitive to alternating
$k$-dependent changes of the symmetry sector and degeneracy of the
ground state sector of $\hat H_{\X, k}$ (note that the exact ground
state symmetry is somewhat relative within the NRG framework, given
an essentially gapless continuum of states of the full system).
The overlap $z_{GG}$ (data marked by triangle) gives the overlap of
the initial and final \emph{ground state} spaces, but is sensitive to
changes in symmetry sector; in particular, for $k\lesssim 28$ it is
nonzero for odd iterations only. The reason why it can be vanishingly
small for certain iterations is, in the present case, that the initial and
final occupancies of the local level differ significantly, as seen
from the values for $\langle n^\I_\imp \rangle$ and $\langle
n^\F_\imp \rangle$ specified in the panel. Therefore, initial and
final ground states can be essentially orthogonal, in the worst case
throughout the entire NRG run. Nonetheless, the AO-exponent is
expected to be well-defined and finite, as reflected in $z_{GK}$.

The AO-measure $z_{KK}$ (data marked by star) is smooth throughout,
and although it is not strictly constrained to the ground state space
at a given iteration, in either initial or final system, it gives the
correct AO exponent, the reason being the underlying energy scale
separation of the NRG. Finally, $z_{\infty,K} = \textrm{Tr}_{\F,k}^K
\{ \hat \rho^G_{\I, \infty} \}$ (data marked by squares) refers to an
AO-measure that calculates the overlap of the ground state space of
an essentially infinite initial system (\ie $k\rightarrow \infty$,
or in practice, the last site of the Wilson chain), with the kept
space at iteration $k$ of the final system. Since the latter
experiences $k$-dependent even-odd differences, whereas the initial
density matrix $\hat \rho^G_{\I, \infty}$ is independent of $k$,
$z_{\infty,K}$ exhibits rather strong $k$-dependent oscillations.
Nevertheless, their envelopes for even and odd iterations separately
decay with the same exponent $\alpha$ as the other AO-measures.

In summary, \Fig{fig:nrg} demonstrates that all AO-measures decay
asymptotically as $e^{\lambda - \alpha k}$, as expected from
\Eq{eq:psi12_wilson}, with the \emph{same} exponent $\alpha$,
independent of the details of the construction. These details only
affect the constant prefactor $\lambda$, which is irrelevant for the
determination of $\dao$.

\subsection{Channel-specific exponents from chains of different lengths}
\label{sec:AO_mu}

Equation (\ref{eq:dao_add}) expresses the exponent $\dao$ of the full
system in terms of the AO exponents $\daomu$ of the individual
channels. This equation is based on the assumption (whose validity
for the models studied here is borne out by the results presented
below) that for distances sufficiently far from the dot, the
asymptotic tail of the ground state wave function factorizes, in
effect, into independent products, one for each channel $\mu$.  This
can be exploited to calculate, in a straightforward fashion, the
individual exponent $\daomu$ for a given channel $\mu$: one simply
constructs a modified Wilson chain which, in effect, is much longer
for channel $\mu$ than for all others. The overlap decay for large
$k$ is then dominated by that channel.

\begin{figure}[tb]
\begin{center}
\includegraphics[width=1\linewidth]{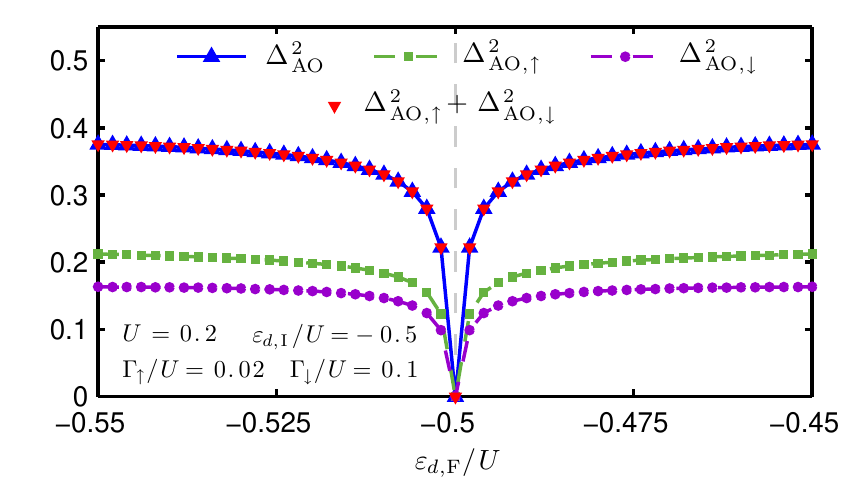}
\end{center}
\caption{(Color online) AO exponents for the standard
    spin-degenerate SIAM with spin-asymmetric hybridization
    [\Eq{eq:SIAM}, with $\mu \in \{\uparrow,\downarrow\}$]
    as functions of $\varepsilon_{\dd,\F}$ (all
    other parameters are fixed as specified in the panel)
 -- The vertical
    dashed line indicates $\varepsilon_{\dd,\I}/U = - 0.5$; at this
    line the initial and final Hamiltonians are identical, hence all
    exponents vanish. The squared AO exponents for the individual channels,
    ${\dao^2}_{, \uparrow}$ (squares) and ${\dao^2}_{,\downarrow}$ (dots),
    were calculated from \Eq{eq:AO_mu}. Their
    sum agrees (with a relative error of less than 1\%) with $\dao^2$
    calculated from \Eq{def.alpha} (down- and upward pointing
    triangles coincide), confirming the validity of the addition rule
    for squared exponents in \Eq{eq:dao_add}.
\label{fig:siam_partial}}
\end{figure}

To be explicit, the strategy is as follows.  First we need to
determine when a Wilson chain is ``sufficiently long'' to capture the
aforementioned factorization of ground state tails. This will be the
case beyond that chain length, say $k_0$, for which the NRG energy
flow diagrams for the kept space excitation spectra of the original
Hamiltonians $\hat H_{\I}$ and $\hat H_{\F}$ are well converged to
their $T=0$ fixed points values.  To calculate $\daomu$, the AO
exponent of channel $\mu$, we then add an artificial term to the
Hamiltonian that in effect depletes the Wilson chain beyond site
$k_0$ for all other channels $\nu \neq \mu$, by drastically raising
the energy cost for occupying these sites.  This term has the form
\begin{equation}
	\label{eq:H_art}
	H_{\rm art} ^{\mu} = C \sum_{\nu \neq \mu} \sum_{k > k_{0}}
	t_{k} \hat{f}^\dagger_{k \nu} \hat{f}^{\pdag}_{k \nu} ,
\end{equation}
with $C \gg 1$. It ensures that occupied sites in the channels $\nu
\neq \mu$ have much larger energy than the original energy scale
$t_{k}$, so that they do not contribute to the low-energy states of
the Hamiltonian. We then calculate a suitable AO-measure (such as
$z_{GK}$) using only $k$-values in the range $k > k_{0}$. From the
exponential decay found in this range, say $\sim e^{- \alpha_\mu k}$,
the channel-specific AO exponent can be extracted,
\cf \Eq{def.alpha},
\begin{eqnarray}
\label{eq:AO_mu}
   \daomu^2  = \frac{4 \alpha_\mu }{ \log \Lambda} \;  .
\end{eqnarray}
This procedure works remarkably well, as illustrated in
\Fig{fig:siam_partial} for the spin-asymmetric single-lead SIAM of
\Eq{eq:SIAM} (with $\NC = 2$, $\mu \in \{ \uparrow, \downarrow\}$).
Indeed, the values for $\daomu$ and $\dao$ displayed in
\Fig{fig:siam_partial} fulfill the addition rule for squared
exponents, \Eq{eq:dao_add}, with a relative error of less than 1\%.

\subsection{Displaced charge}

The displaced charge $\dchargemu$ defined in \Eq{eq:displacedcharge-mu}
can be calculated directly within NRG. However, to properly account
for the contribution from the Fermi sea, $\Delta_{\sea, \mu}$, a
technical difficulty has to be overcome: the Hamiltonians considered
usually obey particle conservation and thus every eigenstate of
$\hat{H}$ is an eigenstate of the total number operator, with an
integer eigenvalue. Consequently, evaluating \Eq{eq:displacedcharge}
over the \emph{full} Wilson chain \emph{always} yields an integer
value for the total $\dchargemu$. This integer, however, does
not correspond to the charge within the large but finite volume
$\Vlarge$ that is evoked in the definition of the displaced
charge.

To obtain the latter, we must consider subchains of shorter length.
Let
\begin{eqnarray}
  \label{eq:charge-up-to-site-k}
   \hat n_{\sea,\mu}^{(k)}
= \sum_{k' = 0}^{k}
\hat{f}_{k'\mu}^{\dagger} \hat{f}_{k'\mu}^{\phantom{\dagger}}
\end{eqnarray}
count the charge from channel $\mu$ sitting on sites 0 to $k$.  These
sites represent, loosely speaking, a volume $\Vlarge^{(k)}$ centered on
the dot, whose size grows exponentially with increasing $k$.  The
contribution from channel $\mu$ of the Fermi sea to the displaced
charge within $\Vlarge^{(k)}$
is
\begin{equation}
	\label{eq:displacedcharge_k}
	\Delta_{\sea,\mu}^{(k)} \equiv
	\left\langle G_{\F}   \right\vert  \hat n_{\sea,\mu}^{(k)}
 \left\vert G_{\F} \right\rangle -
\left\langle G_{\I} \right\vert
  \hat n_{\sea,\mu}^{(k)}
\left\vert G_{\I} \right\rangle ,
\end{equation}
where $|\G_{\I}\rangle$ and $|\G_{\F}\rangle$
are the initial and final ground states
of the \emph{full-length} Wilson chain of length $N$ ($\ge k$).

\begin{figure}[tb]
\begin{center}
\includegraphics[width=1\linewidth]{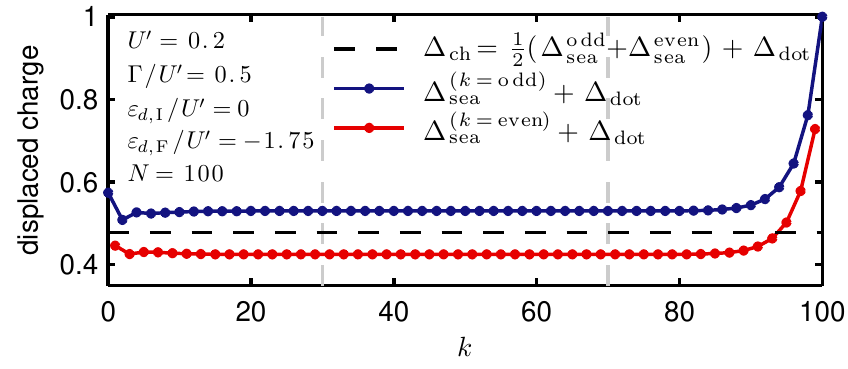}
\end{center}
\caption{(Color online) Determination of $\dcharge$, for the
  interacting resonant level model of \Eq{eq:IRLM}, for a single
  specific set of parameters for $\HI$ and $\HF$, specified in the
  figure legend (the $\varepsilon_{\dd, \F}$-dependence of
  $\dao$ for fixed $\varepsilon_{\dd, \I}$ is analyzed in more detail
  in \Fig{fig:res-level-sweep}.  We obtain $\dcharge$ (dashed
  line) by calculating $\Delta_\sea^{(k)} + \dimp$ and averaging the
  results for even and odd $k$. To reduce the influence of chain's
  boundary regions, we take the average over the region between the
  vertical dashed lines.  }
\label{fig:displacedcharge_wom}
\end{figure}

Figure~\ref{fig:displacedcharge_wom} shows $\Delta_\sea^{(k)}$ for
the spinless IRLM of \Eq{eq:IRLM}, where we dropped the index
$\mu$, since $\NC=1$. $\Delta_\sea^{(k)}$ exhibits even-odd
oscillations between two values, say $\Delta_\sea^\even$ and
$\Delta_\sea^\odd$, but these quickly assume essentially
constant values over a large intermediate range of $k$-values. Near
the very end of the chain they change again rather rapidly, in such a
way that the total displaced charge associated with the full Wilson
chain of length $N$, $\Delta_{\rm ch}^{(N)} = \Delta_\sea^{(N)} +
\Delta_\imp$, is an integer (see \Fig{fig:displacedcharge_wom}),
because the overall ground state has well-defined particle number.
Averaging the even-odd oscillations in the intermediate regime yields
the desired contribution of the Fermi sea to the displaced charge,
$\dsea = \frac{1}{2} (\Delta_\sea^\even + \Delta_\sea^\odd)$.  The
corresponding result for $\dcharge = \dsea + \dimp$ is illustrated by
the black dashed line in \Fig{fig:displacedcharge_wom}.

\section{Results \label{Sec:Results}}

In this section, we present results for the single channel
interacting resonant level model [\Eq{eq:IRLM}], and for single-lead
and two-lead
Anderson impurity models [\Eq{eq:SIAM}]. These examples were chosen
to illustrate that the various ways of calculating AO exponents by
NRG, via $\dao$, $\dphase$ or $\dcharge$, are mutually consistent
with high accuracy, even for rather complex (multi-level, multi-lead)
models with local interactions.  In all cases, the initial and final
Hamiltonians, $\HI$ and $\HF$, differ only in the level position:
$\varepsilon_{d,\I}$ is kept fixed, while $\varepsilon_{d,\F}$ is
swept over a range of values.  This implies different initial and
final dot occupations $n_{d \mu,\X} = \langle G_\X | \hat{n}_{\dd
\mu} | G_\X \rangle$, and hence different local scattering
potentials, causing AO.

\begin{figure}[tb]
\begin{center}
\includegraphics[width=1\linewidth]{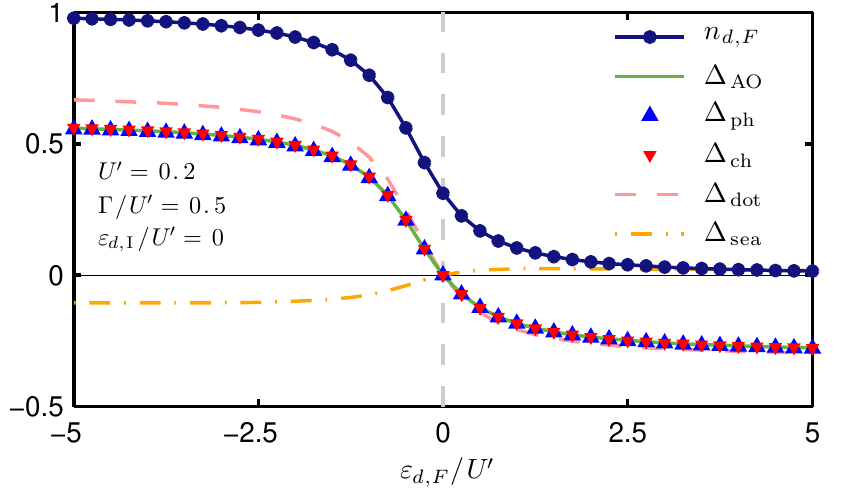}
\end{center}
\caption{(Color online) Verification that $\dao = \dphase =
  \dcharge$ [\Eq{eq:consistency}] for the spinless fermionic
  interacting resonant level model [Eq.~\ref{eq:IRLM}].  All
  quantities are plotted as functions of $\varepsilon_{\dd,\F}$,
  with all other parameters fixed (as specified in the panel).
  The vertical dashed line indicates $\varepsilon_{\dd,\I}/U' =  0$.
  Heavy dots indicate the final
  occupation of the dot, $n_{\dd}$.  The exponent $\dao$ (light solid
  line) agrees well with $\dphase$ and $\dcharge$ (triangles), with
  relative errors of less than $1\%$.  The local and Fermi-sea
  contributions to the displaced charge $\dcharge$ are plotted
  separately, namely $\dimp$ (dashed line) and $\dsea$ (dashed
  dotted). The latter is determined according to the procedure
  illustrated, for $\varepsilon_{d,\F} / U'
  = -1.75$, in \Fig{fig:displacedcharge_wom}.
  \label{fig:res-level-sweep}}
\end{figure}

AO exponents are obtained as described in the previous sections: we
calculate the AO-measure $z_{GK}(k)$ using \Eq{eq:overlap}, obtaining
exponentially decaying behavior (as in \Fig{fig:nrg}). We then
extract $\alpha$ by fitting to $e^{-\alpha k}$ and determine $\dao$
via \Eq{def.alpha}. In the figures below, the resulting $\dao^2$ is
shown as function of $\varepsilon_{\dd\mu, \F}$, together with
$\dcharge^2$, and also $\dphase^2$ in \Fig{fig:res-level-sweep}. The
initial dot level position $\varepsilon_{\dd \mu,\I}$ is indicated by
a vertical dashed line. When $\varepsilon_{\dd \mu,\F}$ crosses this
line, the initial and final Hamiltonians are identical, so that all
AO exponents vanish. To illustrate how the changes in
$\varepsilon_{\dd \mu, \F}$ affect the dot, we also plot the
occupancies $n_{\dd \mu, \F}$ of the dot levels.

\subsection{Interacting resonant-level model}

We begin with a model for which the contribution of the Fermi sea to
the displaced charge is rather important, namely the spinless
fermionic interacting resonant level model [\Eq{eq:IRLM}, $\NC = 1$].
The initial and final Hamiltonians, $\HI^\IRLM$ and $\HF^\IRLM$,
differ only in the level position: the initial one is kept fixed at
$\varepsilon_{d,\I} =0$, while the final one is swept over a range of
values, $\varepsilon_{d,\F} \in [-1,1]$. The results are shown in
\Fig{fig:res-level-sweep}. The final dot occupancy $n_{\dd, \F}$
(heavy dots) varies from $\simeq 1$ to $\simeq 0$, and $\dimp =
n_{\dd, \F} - n_{\dd, \I}$ (dashed line) decreases accordingly, too.
The total displaced charge, $\dcharge = \dimp + \dsea$
(downward-pointing triangles), decreases by a smaller amount, since
the depletion of the dot implies a reduction in the strength of the
local Coulomb repulsion felt by the Fermi sea, and hence an increase
in $\dsea$ (dash-dotted line).  Throughout these changes $\dao$,
$\dphase$ and $\dcharge$ mutually agree with errors of less than
$1\%$, confirming that NRG results comply with \Eq{eq:consistency} to
high accuracy.

\subsection{Single-impurity Anderson model} \label{sec:SIAM}

Next we consider the standard spin-degenerate SIAM for a single lead
[\Eq{eq:SIAM}, $\mu \in \{\uparrow, \downarrow \}$] with
$\varepsilon_{\dd, \mu} = \varepsilon_{\dd}$ and $\Gamma_{\mu} =
\Gamma$. This model exhibits well-known Kondo physics, with a
strongly correlated many-body ground state.

In this model, the dot and Fermi sea affect each other only by
hopping, and there is no direct Coulomb interaction between them ($U'
= 0$). Hence, the contribution of the Fermi sea to the displaced
charge is nearly zero, $\dsea \simeq 0$. Apart from very small
even-odd variations for the first $\sim 35$ bath sites corresponding
to the Kondo scale, the sites of the Wilson chain are half-filled on
average to a good approximation. Therefore $\dsea \ll \dimp$
(explicit numbers are specified in the figure panels; see also
\Fig{fig:nrg}), so that $\dcharge{}_{,\mu}$ in
\Eq{eq:displacedcharge-mu} is dominated by the change of dot
occupation only, \cite{Langreth1966}
\begin{eqnarray}
  \label{eq:deltaN}
  \dcharge^2 \simeq \Delta_\imp^2  \equiv
  \sum_{\mu}
(n_{\dd \mu, \F} - n_{\dd \mu, \I} )^2 \text{.}
\end{eqnarray}
As a consequence, despite the neglect of $\dsea$ in some previous
works involving Anderson impurity models, the Friedel sum rule
($\dphase = \dcharge$) was nevertheless satisfied with
rather good accuracy (typically with errors of a few \%).
However, despite being small, $\dsea$ in practice is on the order of
$|\dsea| \le \Gamma/D$ and thus \emph{finite}. Therefore the
contribution of $\dsea$ to $\dcharge$ will be included throughout,
while also indicating the overall smallness of $\dsea$.
In general, this clearly improves the accuracy of the consistency
checks in \Eq{eq:consistency}, reducing the relative errors to well
below 1\%.

\begin{figure}[tb]
\begin{center}
\includegraphics[width=1\linewidth]{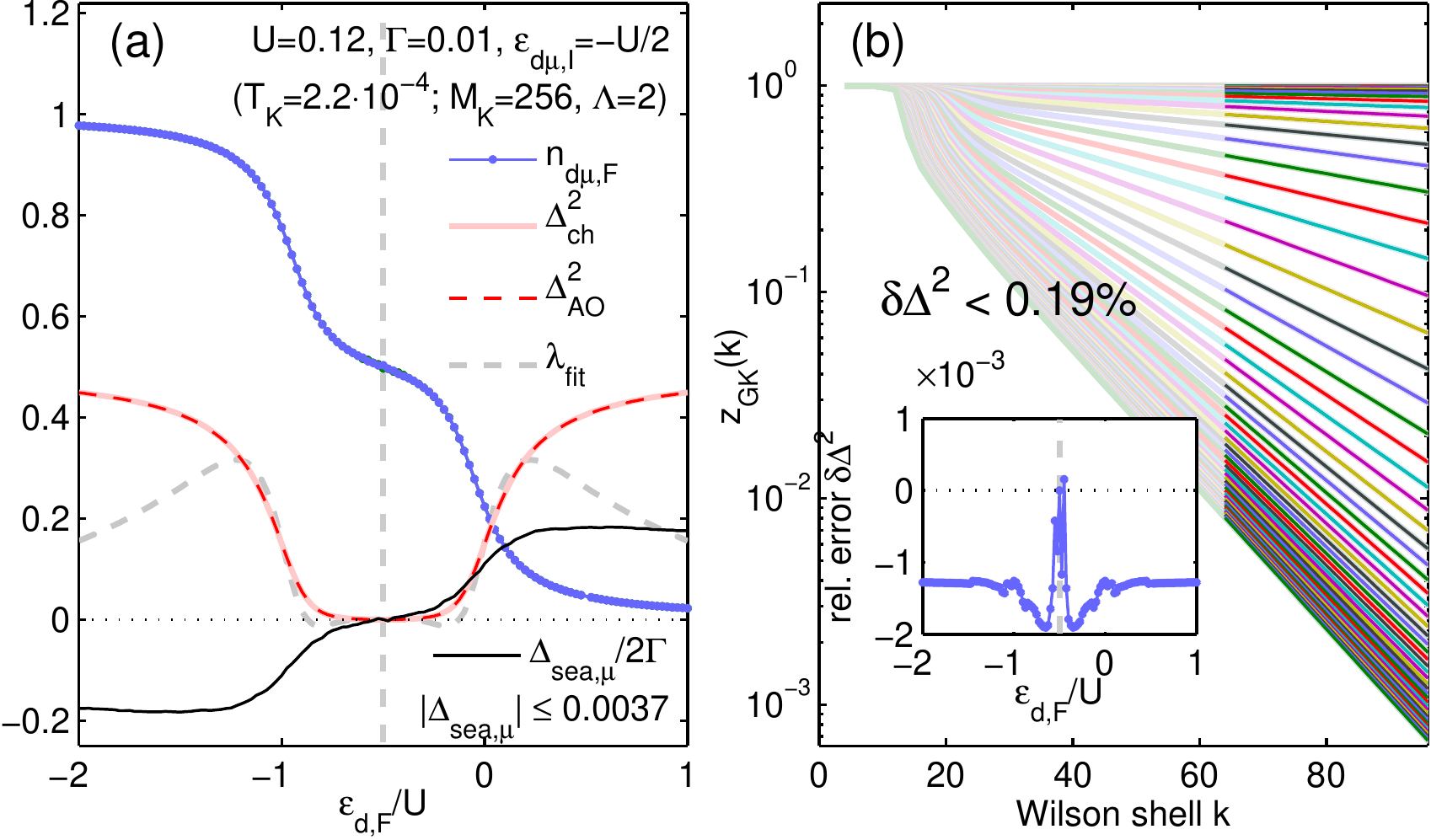}
\end{center}
\caption{
(Color online) Anderson orthogonality for the single-lead,
spin-symmetric SIAM [\Eq{eq:SIAM}, with parameters as specified in
the legend] -- The energy of the $d$-level of the final system
$\varepsilon_{\dd,\F}$ is swept past the Fermi energy of the bath,
while that of the initial reference system is kept fixed in the Kondo
regime at $\varepsilon_{\dd,\I} = - U/2$, indicated by vertical
dashed line in panel (a) and in inset to panel (b).  Panel (a) shows,
as function of $\varepsilon_{\dd, \F}$, the dot occupation per spin
$n_{\dd\mu}$ (dotted solid line), the contribution to the displaced
charge by the Fermi sea, $\dsea_\mu$ (thin black line),
the displaced charge $\dcharge^2$ (light solid line), and the
parameters of the large-$k$ exponential decay $e^{\lambda-\alpha k}$
of $z_{GK}(k)$ as extracted from panel (b), namely $\lambda$ (thick
dashed line) and $\dao$ (dark dashed line), derived from $\alpha$ via
\Eq{def.alpha}. Panel (b) shows the AO-measure $z_{GK}(k) $
in \Eq{eq:overlap} (light lines) for the range of
$\varepsilon_{\dd,\F}$ values used in panel (a). The heavy lines
shown on top for $k \ge 64$ are exponential fits, the results of which
are summarized in panel (a). The inset shows the relative error in
the AO exponents $\delta\Delta^2 \equiv (\dao^2 -
\dcharge^2)/\dcharge^2$, \ie the deviation between the light solid
and dark dashed curve in panel (a); this error is clearly less than
$1\%$ over the full range of $\varepsilon_{\dd}$ analyzed.
\label{fig:SIAM}%
}
\end{figure}

The Anderson orthogonality is analyzed for the SIAM in detail in
\Fig{fig:SIAM}. The initial system is kept fixed at the particle-hole
symmetric point, $\varepsilon_{\dd,\I}=-U/2$ (indicated also by
vertical dashed line in panel a), where the initial ground state is a
Kondo singlet. The final system is swept from double to zero
occupancy by varying $\varepsilon_{\dd, \F}/U$ from $-2$ to 1. The
final ground state is a Kondo singlet in the regime $n_{\dd \mu, \F}
\simeq 1/2$, corresponding to the intermediate shoulder in panel (a).
Panel (b) shows the AO-measure $z_{GK}(k)$ as function of $k$, for a
range of different values of $\edf$. Each curve exhibits clear
exponential decay for large $k$ (as in \Fig{fig:nrg}), of the form
$e^{\lambda - \alpha k}$.  The prefactor, parameterized by $\lambda$,
carries little physical significance, as it also depends on the
specific choice of $z_{\x\x'}$; its dependence on $\edf$ is shown as
thick gray dashed line in panel (a), but it will not be discussed any
further. In contrast, the decay exponent $\alpha$ directly yields the
quantity of physical interest, namely the AO exponent $\dao^2$ via
\Eq{def.alpha}. Panel (a) compares the dependence on $\edf$ of
$\dao^2$ (dashed line) with that of the displaced charge $\dcharge^2$
(light thick line), that was calculated independently from
Eqs.~(\ref{eq:displacedcharge-mu}) and (\ref{eq:dcharge_add}). As
expected from \Eq{eq:consistency}, they agree very well: the relative
difference between the two exponents $\dao^2$ and $\dcharge^2$ is
clearly below $1\%$ throughout the entire parameter sweep, as shown
in the inset of \Fig{fig:SIAM}(b).

The contribution of the Fermi sea to the displaced charge is close to
negligible, yet finite throughout (black line in panel a).
Overall, $\dsea \lesssim 0.0037$, as indicated in \Eq{eq:deltaN}.
Nevertheless, by including it when calculating $\dcharge$, the
relative error $\delta \Delta^2$ is systematically reduced from a few
percent to well below 1\% throughout, thus underlining its
importance.

\begin{figure}[tb]
\begin{center}
\includegraphics[width=0.9\linewidth]{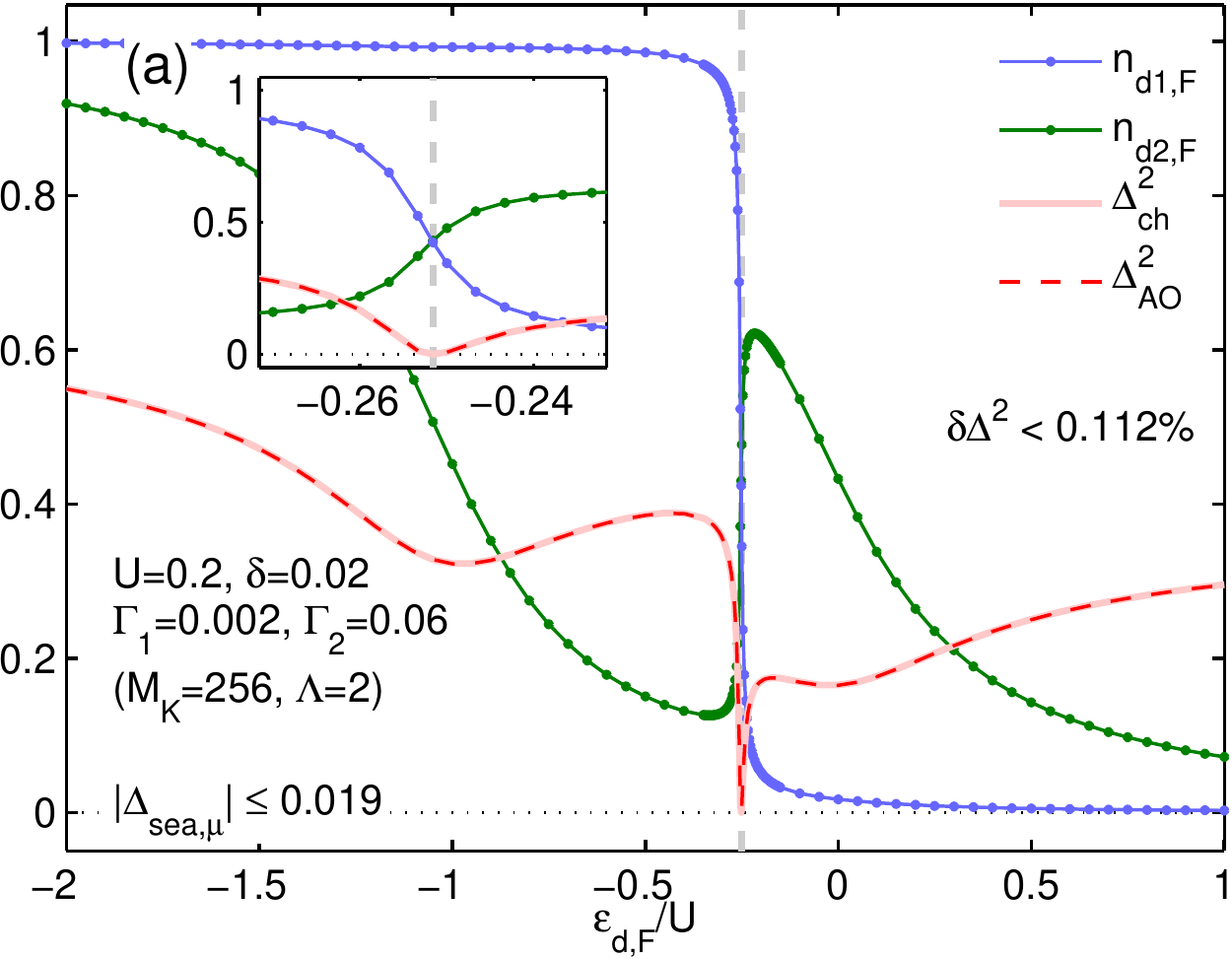}\\[1ex]
\includegraphics[width=0.9\linewidth]{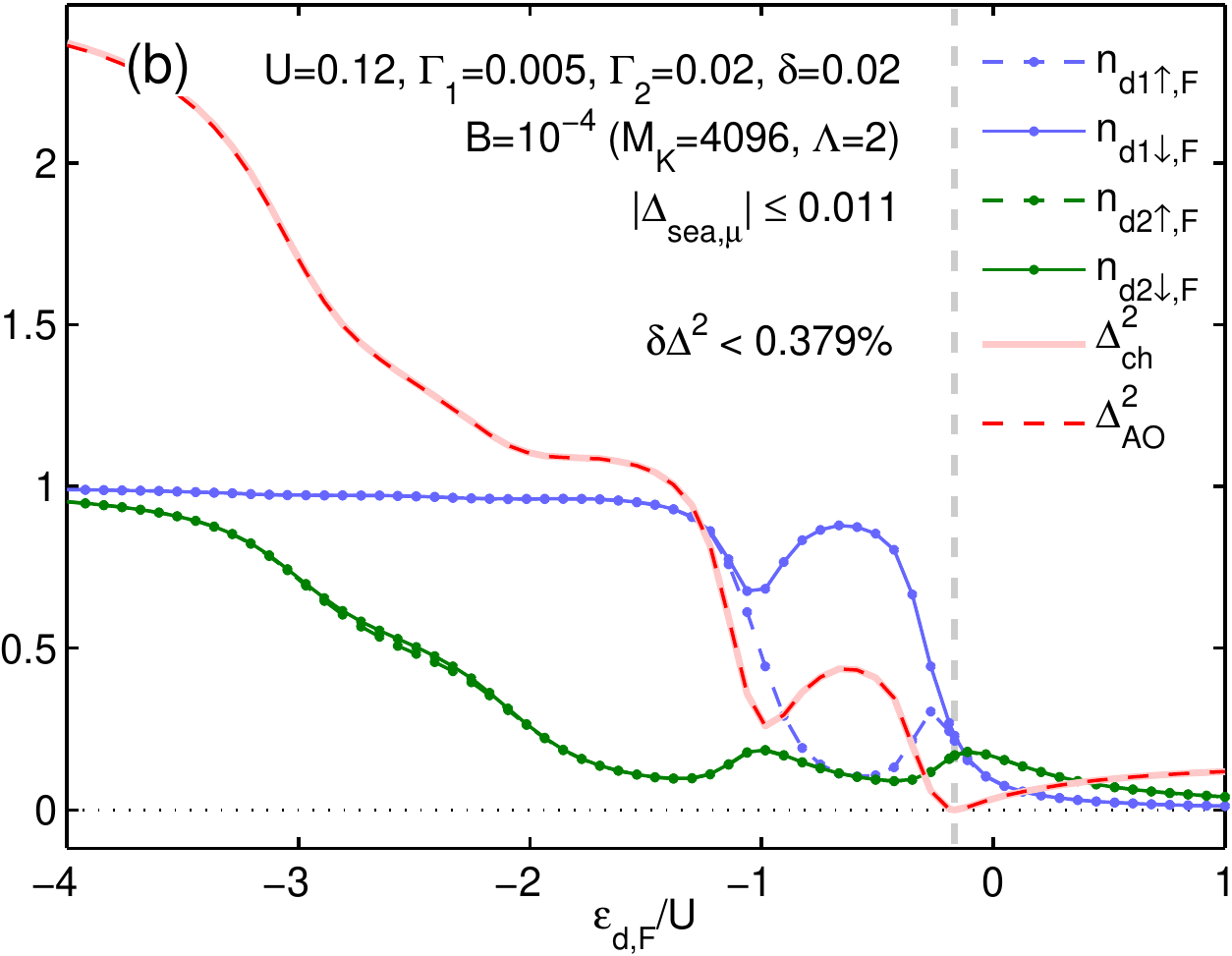}
\end{center}
\caption{
(Color online) Anderson orthogonality for a spinless (panel a) and
spin-full (panel b) two-lead SIAM, with dot levels of unequal width
and a split level structure as defined in \Eq{eq:levelsplittings}
(all relevant model parameters are specified in the legends) -- In
both cases, the higher level~2 is broader than the lower level~1
$(\Gamma_2 > \Gamma_1)$, leading to population switching as function
of the average final level energy $\varepsilon_{\dd, \F}$. The fixed
value of  $\varepsilon_{\dd,\I}$ is indicated by the vertical dashed
line. The inset to panel (a) shows a zoom into the switching
region, clearly demonstrating that population switching occurs
smoothly.
For panel (b), a finite magnetic field $B$ causes a splitting between
spin-up and spin-down levels, resulting in a more complex switching
pattern. In both panels $\dao^2$ and $\dcharge$ agree very well
throughout the sweep, with a relative error $\delta\Delta^2$ well
below $1\%$.}
\label{fig:SIAM_2LS}
\end{figure}

\subsection{Multiple Channels and Population switching}

Figure \ref{fig:SIAM_2LS} analyzes AO for lead-asymmetric two-level,
two-lead SIAM models, with Hamiltonians of the form \Eq{eq:SIAM}
(explicit model parameters are specified in the panels).
Panel (a) considers a spinless case ($\NC=2$, $\mu = j \in \{1,2\}$),
whose dot levels have mean energy $\varepsilon_{\dd}$ at fixed
splitting $\delta$,
\begin{subequations}
\label{eq:levelsplittings}
\begin{eqnarray}
\label{eq:i-splitting}
\varepsilon_{\dd 1 } = \varepsilon_{\dd} - \delta/2 , \quad
\varepsilon_{\dd 2 } = \varepsilon_{\dd} + \delta/2 .
\end{eqnarray}
Panel (b) considers a spinfull case, ($\NC=4$, $\mu = (j\sigma)$ with
$j\in \{1,2\}$, $\sigma \in \{ \uparrow, \downarrow\}$), where both
the lower and upper levels have an additional (small) spin splitting
$B \ll \delta$,
\begin{eqnarray}
\label{eq:sigma-j-splitting}
\varepsilon_{\dd j\uparrow } = \varepsilon_{\dd j} + B/2 , \quad
\varepsilon_{\dd j\uparrow } = \varepsilon_{\dd j} - B/2 \; .
\end{eqnarray}
\end{subequations}
Charge is conserved in each of the $\NC$ channels, since these only
interact through the interaction on the dot. In both models, the
upper level~2 is taken to be broader than the lower level~1,
$\Gamma_{2} > \Gamma_1$ (for detailed parameters, see figure
legends). As a consequence,
\cite{Silvestrov2000,Golosov2006,Karrasch2007PRL,Goldstein2010} these
models exhibit population switching: when $\varepsilon_{\dd, \F}$ is
lowered (while all other parameters are kept fixed), the final state
occupancies of upper and lower levels cross, as seen in both panels
of \Fig{fig:SIAM_2LS}.

Consider first the spinless case in Figure \ref{fig:SIAM_2LS}(a). The
broader level~2 shows larger occupancy for large positive
$\varepsilon_{\dd, \F}$. However, once the narrower level~1 drops
sufficiently far below the Fermi energy of the bath as
$\varepsilon_{\dd,\F}$ is lowered, it becomes energetically favorable
to fill level~1, while the Coulomb interaction will cause the level~2
to be emptied. At the switching point, occupations can change
extremely fast, yet they do so smoothly, as shown in the zoom in the
inset to panel (a).

Similar behaviour is seen for the spinfull case in Figure
\ref{fig:SIAM_2LS}(b), though the filling pattern is more complex,
due to the nonzero applied finite magnetic field $B$ (parameters are
listed in the legend).  The occupations $n_{\dd 1 \sigma}$ of the
narrower level~1 show a strong spin asymmetry, since the magnetic
field is comparable, in order of magnitude, to the level width ($B =
\Gamma_1/2$). This asymmetry affects the broader level~2, which fills
more slowly as $\varepsilon_d$ is lowered. Due to the larger width of
level~2, the asymmetry in its spin-dependent occupancies is
significantly weaker. As in panel (a), population switching between
the two levels occurs: as the narrower level~1 becomes filled, the
broader level~2 gets depleted.

The details of population switching, complicated as they are
(extremely rapid in panel (a); involving four channels in panel (b)),
are not main point of \Fig{fig:SIAM_2LS}. Instead, its central
message is that despite the complexity of the switching pattern, the
relation $\dao^2 = \dcharge^2$ is satisfied with great accuracy
throughout the sweep (compare light thick and dark dashed lines).
Moreover, since $\dcharge$ was calculated by adding the contributions
from separate channels according to \Eq{eq:dcharge_add}, this also
confirms the additive character of AO exponents for separate
channels.

As was the case for the single-channel SIAM discussed in
Sec.~\ref{sec:SIAM} above, a direct interaction between dot and Fermi
sea is not present in either of the models considered here $(U'=0)$.
Consequently, the displaced charge $\dcharge$ is again dominated by
$\dimp$, with $\dsea \ll \dimp$ (\cf Eq.~\ref{eq:deltaN}).
Specifically, for the spinless or spinfull models, we find  $\dsea <
0.019$ or $0.011$, respectively, for the entire sweep.

  \section{Summary and Outlook \label{Sec:Summary}}

In summary, we have shown that NRG offers a straightforward,
systematic and self-contained way for studying Anderson
orthogonality, and illustrated this for several interacting quantum
impurity models.  The central idea of our work is to exploit the fact
that NRG allows the size-dependence of an impurity model to be
studied, in the thermodynamic limit of $N\to \infty$, by simply
studying the dependence on Wilson chain length $k$. Three different
ways of calculating AO exponents have been explored, using
wave-function overlaps ($\dao$), changes in phase shift at the Fermi
surface ($\dphase$), and changes in displaced charge ($\dcharge$).
The main novelty in this paper lies in the first of these, involving
a direct calculation of the overlap of the initial and final ground
states themselves. This offers a straightforward and convenient way
for extracting the overall exponent $\dao$. Moreover, if desired, it
can also be used to calculate the exponents ${\dao}_{,\mu}$
associated with individual channels, by constructing a Wilson chain
that is longer for channel $\mu$ than for the others. We have also
refined the calculation of $\dcharge$, by showing how the
contribution $\dsea$ of the Fermi sea to the displaced charge can be
taken into account in a systematic fashion.

The resulting exponents $\dao$, $\dphase$ and $\dcharge$ agree
extraordinarily well, with relative errors of less than 1\%. This can
be achieved using a remarkably small number of kept states $M_K$. For
example, for the spinfull SIAM analyzed below, a better than 5\%
agreement can be obtained already for $M_K \ge 32$. (For comparison,
typically $M_K = 250$ is required to obtain an accurate description
of the Kondo resonance of the $d$-level spectral function in the
local moment regime of this model.)

Our analysis has been performed on models exhibiting Fermi liquid
statistics at low temperatures. As an outlook, it would be interesting
to explore to what extent the non-Fermi liquid nature of a model would
change AO scaling properties, an example being the symmetric spinful
two-channel Kondo model.

Finally, we note that non-equilibrium simulations of quantum impurity
models in the time-domain in response to quantum quenches are a
highly interesting topic for studying AO physics in the time domain.
The tools to do so using NRG have become accessible only rather recently.
\cite{Anders2005,Anders2006,Wb07,Tureci2011} One considers a sudden
change in some local term in the Hamiltonian and studies the
subsequent time-evolution, characterized, for example, by the
quantity $\langle G_\I | e^{-i \HF t} | G_\I \rangle$.  Its numerical
evaluation requires the calculation of overlaps of eigenstates of
$\HI$ and $\HF$. The quantity of present interest, $|\langle G_\I |
G_\F \rangle|$, is simply a particular example of such an overlap. As
a consequence, the long-time decay of $\langle G_\I | e^{-i \HF t} |
G_\I \rangle$ is often governed by $\dao$, too,
\cite{Nozieres1969,Schotte1969} showing power-law decay in time with
an exponent depending on $\dao$. This will be elaborated in a
separate publication.\cite{Muender2011b}

\begin{acknowledgements}
We thank G. Zar\'and for an inspiring discussion that provided the
seed for this work several years ago, and Y. Gefen for encouragement
to pursue a systematic study of Anderson orthogonality. This work
received support from the DFG (SFB 631, De-730/3-2, De-730/4-2,
WE\-4819/1-1, SFB-TR12), and in part from the NSF under Grant No.
PHY05-51164. Financial support by the Excellence Cluster
``Nanosystems Initiative Munich (NIM)'' is gratefully acknowledged.
\end{acknowledgements}



\end{document}